Three Universal Distribution Functions for Native Proteins with Harmonic Interactions

Burak Erman

Chemical and Biological Engineering Department, Koc University

Istanbul, Turkey

email: berman@ku.edu.tr

We used statistical thermodynamics of conformational fluctuations and the elements of algebraic graph theory together with data from 2000 protein crystal structures, and showed that folded native proteins with harmonic interactions exhibit distribution functions each of which appear to be universal across all proteins. The three universal distributions are: (i) the eigenvalue spectrum of the protein graph Laplacian, (ii) the B-factor distribution of residues, and (iii) the vibrational entropy difference per residue between the unfolded and the folded states. The three distributions, which look independent of each other at first sight, are strongly associated with the Rouse chain model of a polymer as the unfolded protein. We treat the folded protein as the strongly perturbed state of the Rouse chain. We explain the underlying factors controlling the three distributions and discuss the differences from those of randomly folded structures.

INTRODUCTION

Following the seminal works of Tirion [1] and ben-Avraham [2], the area of research for describing proteins with simple potentials and determining universal features of vibrational spectra in globular proteins received widespread attention [3-11]. These studies, which are rooted on earlier work on normal mode analysis [12-14] focused mainly on the low frequency region of the vibrational spectrum, and tested the hypothesis of universality on a limited number of crystal structures. Only very recently, using an all atom CHARMM force-field and a data set of 135 protein crystal structures, Na et. al, [15] showed that the vibrational spectrum of globular proteins is universal over the full range of frequencies. In the present letter, using 2000 protein crystal structures, we show that the universality of the full relaxation spectrum goes well beyond the details of a force field and is indeed a result of the graph-model of proteins with simple harmonic interactions between residues. Using the same data set, we also show, for the first time, that when properly scaled, the fluctuations of residues, the B-factors, are not independent but obey a universal distribution that maximizes the Shannon entropy. Thirdly, we show that the decrease of vibrational entropy per residue upon folding of the proteins into the globular native state obeys a universal distribution that is a sharply peaked Gaussian, a feature that is surprisingly different than that for randomly folded structures where the distribution is highly dispersed, larger entropy decrease being favored. We show that these three apparently unrelated universal features are indeed strongly associated with different level perturbations of the unfolded chain. These new findings give us a simple and clear picture of folded proteins and a better fundamental understanding of why proteins behave the way they do.

A newly synthesized protein emerging from the ribosome resembles a Rouse chain [16] exhibiting large scale spatial fluctuations under short range interactions among residues close to each other along the chain. Nearest neighbor residue interactions ensure chain connectivity. $i, i \pm k$ type interactions with $3 \le k \le 4$ correspond to short range interactions in a helical structure. Long range interactions are between residue pairs widely separated along the chain. The magnitude of fluctuations of the Rouse chain with no long range interactions is quadratic with respect to the central residue of the chain [16]. Different values of k for short range interactions have the effect of changing the curvature of the parabola only. The protein 1LTSc.pdb,

shown in Figure 1a has an extended helical structure with only short range interactions up to $k=4$. Due to its extended structure, it has no long range interactions, and therefore comes closest to the Rouse chain. In Figure 1b, the experimental B-factors of 1LTS and of a Rouse chain with interactions up to $i \pm 4$ are compared.

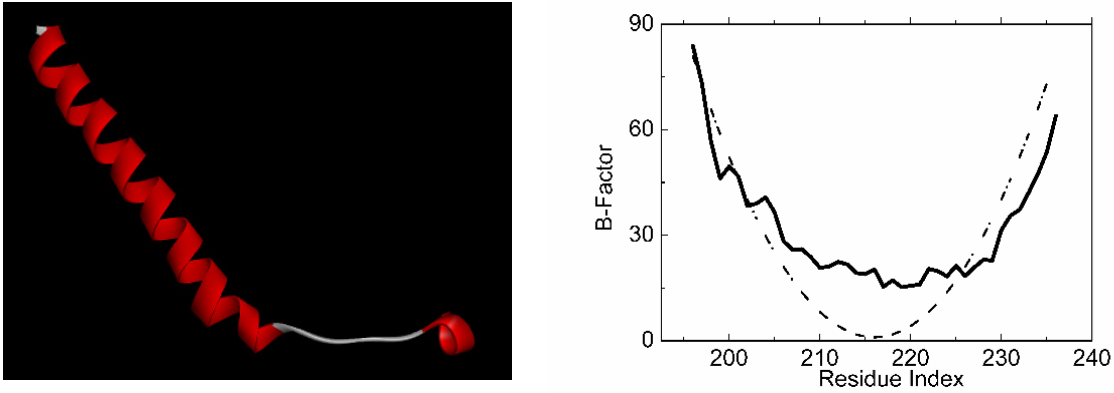

FIG 1. (a) (Color online) ribbon diagram of 1LTS.pdb, (b) B-factor values of 1LTS.pdb (thick solid curve) and the Rouse chain (dashed curve).

The B-factor of a residue or the Debye-Waller factor of each residue is listed in Protein Data Bank (PDB) files and is a direct measure of the mean square fluctuations, $\left\langle (\Delta R)^2 \right\rangle$, according to the relation $\left\langle (\Delta R_i)^2 \right\rangle = \frac{3}{8\pi^2} B_i$, where $\Delta R_i$ denotes the fluctuation in the position vector of the ith residue. In the absence of long range interactions the fluctuation profile of a protein is well approximated by a parabola, the tails fluctuating the most, as is the case for the real protein 1LTSc. One may intuitively consider the folding of a protein as a process starting with the Rouse state and progressing into denser states during which the number of long range interactions increases and fluctuations decrease until the system reaches the folded state. In this final state, the fluctuations that remain, the Protein Data Bank B-factor profiles, are those that are allowed to the residues subject to the constraints imposed by the equilibrium structure. It is intuitively useful to regard the B-factor profiles of native proteins as strongly perturbed Rouse curves. Under thermal energy, it is obvious that the native state fluctuations will have a tendency to increase indefinitely towards those of the Rouse state, but will not be able to do so due to the constraints imposed by the native equilibrium structure that is trapped at a minimum energy state, i.e., the equilibrium state will maximize the vibrational entropy of residues constrained by the minimum energy equilibrium structure.

## THEORY AND THE MODEL

The coarse graining approximation is adopted in this study, where atoms of a residue are collapsed on the alpha carbon, and the positions of the alpha carbons are the only variables. The correlation of instantaneous fluctuations of two residues i and j is related to the equilibrium value $R_i$ of residue i and the equilibrium force $F_j$ on residue j by the Onsager relation [17]

$$\left\langle \Delta R_i \Delta R_j \right\rangle = k_B T \frac{\partial R_i}{\partial F_j} \qquad (1)$$

Here, $k_B$ is the Boltzmann constant, $T$ is the absolute temperature. This equation is general, the right hand side being obtained from an equation of state, $F_i = f(R_1, R_2, ..., R_n)$. In the harmonic approximation, the equation of state is a linear one based on neighbor interactions

$$F_i = \Gamma_{ij} R_j \tag{2}$$

where, $\Gamma_{ij}$ is the spring constant matrix,

$$\Gamma_{ij} = \begin{cases} d_i & \text{if } i = j \\ -k_{ij} & \text{if } i \text{ and } j \text{ are adjacent} \\ 0 & \text{otherwise} \end{cases} \tag{3}$$

Here, two residues are adjacent if one lies within the first coordination shell of the other. $k_{ij}$ is the spring constant of the spring that connects i and j. $d_i$ is the sum of all spring constants between i and all of its neighbors. In graph theory jargon, a protein P is a graph, the ith residue is the ith vertex of the graph, and an interaction between two residues is an edge of the graph. The protein with harmonic interactions corresponds to a weighted graph Laplacian where $k_{ij}$'s are the weights of edges. Earlier work showed that a single value for the spring constant works almost perfectly [1,9]. Following this simplification all spring constants are taken equal, and equal to unity, the exact value of the spring constant not being essential here. In this case, Eq. 3 becomes the Laplacian of the protein, $d_i$ becomes the degree of the ith vertex. Substituting Eq. 2 into Eq. 1 leads to the basic equation of the Gaussian Network Model, GNM [9]

$$\langle \Delta R_i \Delta R_j \rangle = kT \Gamma_{ij}^{-1} \tag{4}$$

Despite the simplifications underlying the model, predictions based on the GNM are often as accurate as those of detailed all atom molecular dynamics studies. The underlying cause of this seemingly contradictory statement is the impact of the universal properties of proteins that lack all detail. The main purpose of this study is to extract three such universal properties based on a set of 2000 non-homologous protein crystal structures. The list of the proteins used is given in Supplementary Material.

Let $T$ denote the diagonal matrix with the (i,i)'th entry having value $d_i$ and with the convention that $T^{-1}_{ii} = 0$ if $d_i = 0$. The reduced Laplacian $\gamma_{ij}$ of $\Gamma_{ij}$ is defined as the matrix $\gamma = T^{-1/2} \Gamma T^{-1/2}$, which reads as [18]

$$\gamma_{ij} = \begin{cases} 1 & \text{if } i = j \text{ and } d_j \neq 0 \\ \dfrac{-1}{\sqrt{d_i d_j}} & \text{if } i \text{ and } j \text{ are adjacent} \\ 0 & \text{otherwise} \end{cases} \tag{5}$$

The eigenvalues of the reduced Laplacian vary between 0 and 2 [18]. The Laplacian $\Gamma_{ij}$ may be expressed in spectral form as

$$\Gamma_{ij} = \sum_k v_{ik} v_{jk} \lambda_k \tag{6}$$

where $\lambda_k$ is the k'th eigenvalue and $v_{ik}$ is the eigenvector corresponding to the kth mode. The ij'th element of the inverse of $\Gamma_{ij}$ is

$$\left(\Gamma^{-1}\right)_{ij} = \frac{1}{k_B T}\langle \Delta R_i \Delta R_j \rangle = \sum_k v_{ik} v_{jk} \lambda_k^{-1} \qquad (7)$$

The matrix $\Gamma$ is singular and one of the eigenvalues is zero with the corresponding eigenvector equal to $\mathbf{1}$, i.e., a vector with all entries equal to unity.

Let $g$ denote an arbitrary function which assigns to each vertex of the graph a real value $g_i$. The Laplacian $\gamma_{ij}$ may be regarded as an operator on $g$ which satisfies [18]

$$\gamma g_i = \frac{1}{\sqrt{d_i}} \sum_{\substack{j \\ i\sim j}} \left( \frac{g_i}{\sqrt{d_i}} - \frac{g_j}{\sqrt{d_j}} \right) \qquad (8)$$

where, $\sum_{\substack{j \\ i\sim j}}$ denotes summation over j for which i is a neighbor of j.

A new function $f_i$ may be defined at every vertex of the graph as $f = T^{-1/2} g$. Then, the second smallest eigenvalue $\lambda_{\gamma 2}$ of $\gamma$ is related to $f$ by the Rayleigh quotient [18]

$$\lambda_{\gamma 2} \leq \frac{\sum_{j\sim i}(f_i - f_j)^2}{\sum_i f_i^2 d_i} \qquad (9)$$

The equality holds if $f \perp T^{1/2}\mathbf{1}$. In this case, $f$ is called a harmonic eigenfunction of $\gamma$. The general case obtains if we can find a function $f$ orthogonal to the subspace $P_{k-1}$ generated by the k-1 harmonic eigenfunctions:

$$\lambda_k = \inf_{f \perp P_{k-1}} \frac{\sum_{j\sim i}(f_i - f_j)^2}{\sum_i f_i^2 d_i} \qquad (10)$$

where, inf denotes infimum. Equation 10 can be used, as shown below, in the analysis of correlations in proteins since the eigenvectors of $\Gamma$ and $\Gamma^{-1}$ are identical.

**_The Laplacian spectrum:_** In Figure 2, the frequency density of eigenvalues of the reduced Laplacian obtained from the set of 2000 proteins is shown by the small dots, about 401000 in number for the 2000 proteins. Large solid circles are the averages taken at each eigenvalue interval. The solid line is the result for random graphs obtained [19] by connecting two vertices by an edge with a probability P= 0.09. This probability gives an average vertex degree of 8.5 which is the average value of vertex degrees of globular proteins. The curve is an average of 10,000 chains whose lengths varied between 50 and 350.

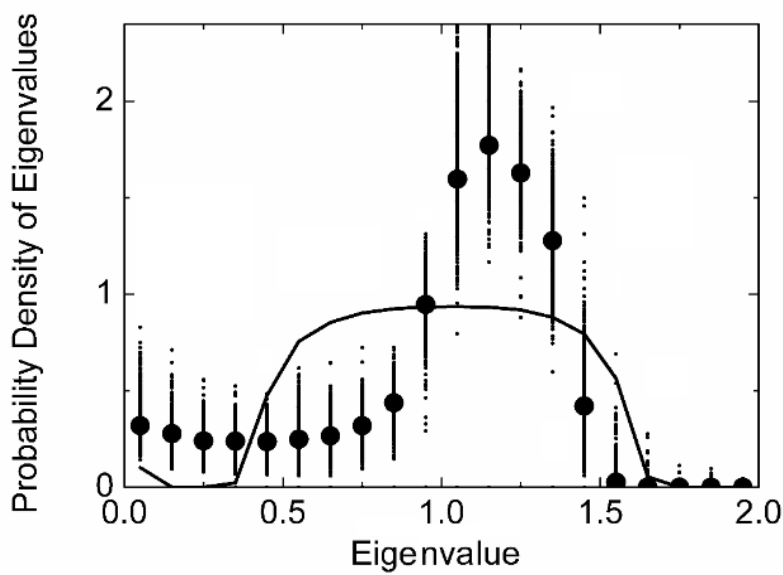

FIG 2. Distribution of eigenvalues of 2000 non-homologous proteins and of 10,000 random networks.

Considering the simplicity of the potential used which contains no adjustable parameters, the large extent of the outliers in the figure is within a plausible range. Recently, Na et. al., [15] used an all-atom nonlinear potential which led to a comparable level of outliers. In Figure 2, for eigenvalues less than unity, which corresponds to the fast motion regime, the probability density of eigenvalues for native proteins is small and approximately constant. The maximum of the eigenvalues of native proteins lies between 1.0 and 1.5, corresponding to the slow modes region. Above 1.5 the eigenvalues are close to zero. The spectrum for the random case has a distribution symmetric around 1.0.

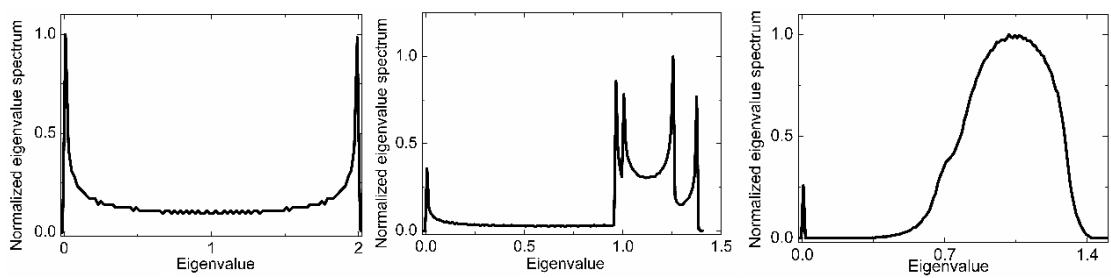

FIG 3. Normalized eigenvalue spectra of (a) Rouse chain with nearest neighbor interaction i and i+1, (b) Rouse chain with short range interactions up to i+4, (c) Rouse chain of part b plus random interactions with P=0.09.

In Figure 3, the spectra for the Rouse chain of length varying randomly between 50 and 350 residues are presented. In part a, the Rouse chain with nearest neighbor interactions $i, i \pm 1$ only is shown. The spectrum is symmetric with two maxima at the extremities and is flat in between. In part b, a Rouse chain in the presence of short range interactions, $1 \leq k \leq 4$, but no long range interactions is shown. The spectrum surprisingly started to resemble the native protein spectrum. The symmetry of the spectrum is lost. The region between 0 and 1 is now low amplitude and flat, and several maxima appeared in the region between 1 and 1.5. In part c, random long range interactions with P=0.09 are superimposed on the chain of part b. The shape now has all the features of native protein eigenvalue spectrum of Figure 2. In conclusion, the most important contribution to the spectrum of real as well as random proteins comes from short range interactions $i, i \pm k$, $k > 1$, which not only modify the fast motions range but also shift the slow motions peak. Random contacts or selective contacts as in real proteins appear to be of secondary effect in determining the shape of the spectrum. The spectrum obtained by harmonic interactions is in remarkable qualitative resemblance to all atom CHARMM potential results [15].

***The distribution of B-factors:*** Next, we show that the B-factors are not independent of each other in a protein. For simplicity, we use $B_i = \langle (\Delta R_i)^2 \rangle$, leaving out the factor $\frac{8\pi^2}{3}$ which may be incorporated into the formulation if quantitative agreement is of interest. In spectral representation, B-factors in the k'th mode are written as

$$B_i(k) = \frac{u_i^2(k)}{\lambda_k} \tag{11}$$

where, $u_i(k)$ is the ith component of the k'th eigenvector corresponding to the eigenvalue $\lambda_k$. For a given mode $k$, $u_i(k)$ and therefore $\frac{u_i(k)}{\sqrt{\lambda_k}} = \sqrt{B_i(k)}$ are eigenvectors of $\Gamma$. We now choose $g_i = \frac{u_i(k)}{\sqrt{\lambda_k}} = \sqrt{B_i(k)}$. Consequently, $f_i = \sqrt{\frac{B_i(k)}{d_i}}$ becomes a harmonic eigenfunction of the reduced Laplacian. Substituting this expression into Eq. 10 and using the relation $\sum_i B_i(k) = \frac{\sum_i u_i^2(k)}{\lambda_k} \equiv \frac{1}{\lambda_k}$ leads to a relationship between $B_i(k)$ for each k as:

$$\sum_{j \sim i} \left[ \frac{B_i(k)}{d_i} - 2\sqrt{\frac{B_i(k) B_j(k)}{d_i d_j}} - \frac{B_j(k)}{d_j} \right] = 1 \tag{12}$$

showing that the B-factors of the protein are not independent but obey the relationship imposed by Eq. 12 for each mode, k. Equation 12 may be called the Rayleigh relation for B-factors because it is based on the Rayleigh quotient of algebraic graph theory [18].

The energy of the ith residue in the k'th mode is [5]

$$E_i(k) = u_i^2(k) = \lambda_k B_i(k) \tag{13}$$

Summing over all modes and all residues gives the total energy $E_{total}$ of the protein:

$$E_{total} = \sum_i \sum_k E_i(k) = \sum_i \sum_k \lambda_k B_i(k) = \sum_i \sum_k u_i^2(k) \equiv N - 1 \tag{14}$$

Denoting the distribution of B-factors in the k'th mode by $w_i(k)$, the variation of the Shannon entropy may now be equated to zero subject to constant energy

$$\delta \left\{ \sum_i w_i(k) \ln w_i(k) + \eta \sum_i \left[ w_i(k) \lambda_k B_i(k) - E_i(k) \right] \right\} = 0 \quad for\ k = 2,...,N \tag{15}$$

where $\eta$ is the Lagrange multiplier. Solution of Eq. 15 leads to the distribution of B values for each mode as

$$w_i(k) = Ce^{-\lambda_k B_i(k)} \quad for\ k = 2,...,N \tag{16}$$

Since the modes are independent, the distribution $w(B_i)$ over all of the modes will be the convolution of each modal distribution which will be the Gamma distribution

$$w(B_i) = \alpha B_i^{p-1} e^{-B_i/\tau} \tag{17}$$

where $\alpha$, $p$ and $\tau$ are parameters. In Figure 4, results of experimental B-factor distributions from the set of 2000 proteins are presented by circles. The abscissa in Figure 4 is obtained by subtracting the minimum B-factor and dividing the difference with the maximum B-factor for each protein. The ordinate values are obtained by counting the frequency of the B values in bins of 0.01 wide intervals between 0 and 1. The distribution is obtained from a total of 403755 B-factor values for 2000 proteins. The experimental points are normalized so that the largest value of the ordinate is unity. The solid curve is obtained from Eq. 17 with $\alpha$ =33.98, p = 2 and and $\tau$ =0.08.

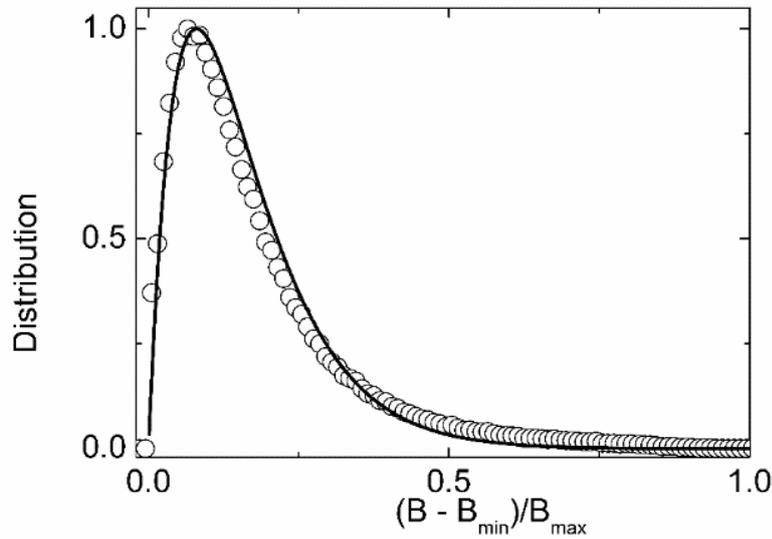

FIG 4. Distribution of B-factors. Circles are from experimental data, the solid curve is from Eq. 17.

Figure 4 shows that the B-factors obey the Gamma distribution. The actual, unscaled distribution of B-factor values for a given protein may be obtained from the curve of Figure 4 if $B_{min}$ and $B_{max}$ value of the protein are known. Consequently, the entropy and the free energy will be determined which is not pursued further here. It is worth noting that the imposition of the distribution precludes independent changes in the B-factors. A change in the B-factor of one residue should be accompanied by other changes to preserve the distribution. Several phenomena in proteins, such as correlated mutations in coevolution support this basic property of dependence.

*Entropy change upon folding*: The change in the vibrational entropy of proteins per residue in going from the Rouse state to the folded state is given as [5]

$$\frac{\Delta S}{Nk} = -\sum \ln \frac{\lambda_i}{\lambda_{iR}} \tag{18}$$

where $\lambda_{iR}$ is the ith eigenvalue of the Rouse chain. A factor of 3/2, present in Reference [5] is not shown in Eq. 18 for simplicity of presentation. Results of calculations for the distribution of entropy for the 2000 proteins are shown in Figure 5 by the filled circles. Empty circles show the results for randomly folded proteins calculated over 10,000 chains of random lengths between 50 and 350. The results are normalized to unity for both sets. The solid lines are the best fitting curves, a sharply peaked Gaussian for the real proteins and an exponential for the random folders. The distribution for real proteins indicates that the majority of them have approximately the same entropy decrease upon folding. For random folders, exponentially dispersed points indicate that the majority chains require large entropy decreases for folding.

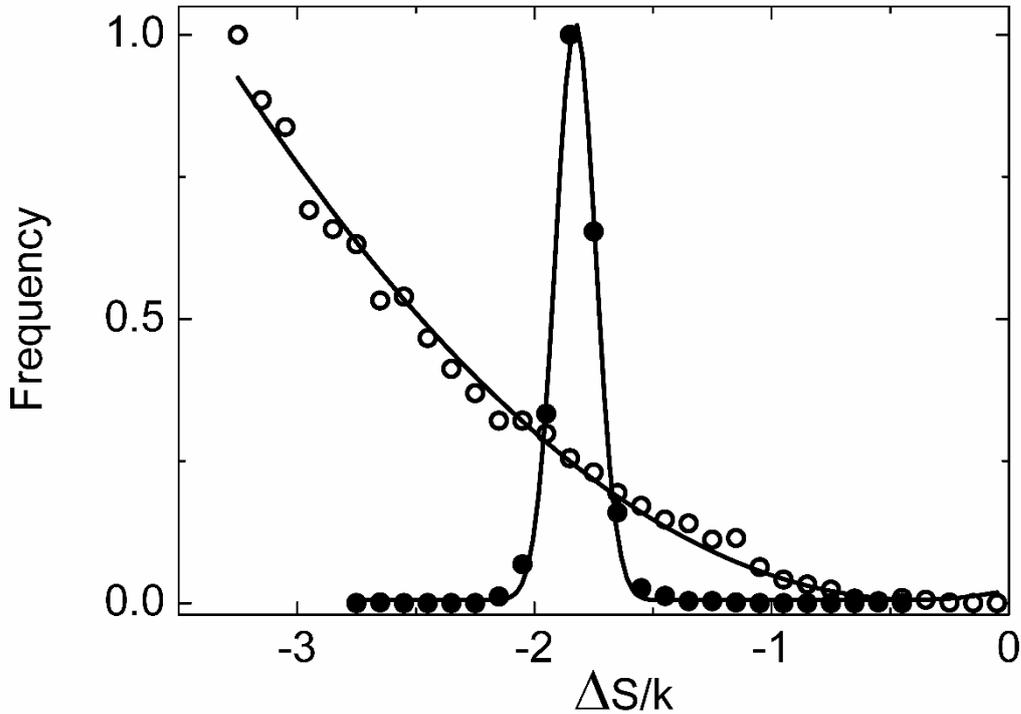

FIG 5. Vibrational entropy change in going from the Rouse state to the folded state. Filled points are for the 2000 native proteins, empty circles are for 10,000 randomly folded proteins. Lines are best fitting curves, a Gaussian for real proteins and an exponential relation for random folders.

CONCLUSION

A closer correspondence of the present model to real proteins awaits improvements in passing from Eq. 1 to Eq. 2 by introducing anharmonicities which may be essential for some problems for proteins. This is clearly demonstrated in the recent work of Na et. al. [15]. However, for the majority of problems of interest, the harmonic approximation serves as a satisfactory approximation. Once the harmonic approximation is adopted and the protein problem is transformed into the graph Laplacian, a whole new vista is opened that may be extensively exploited with the tools of algebraic graph theory. Among many possibilities offered by algebraic graph theory, the use of the Laplacian operator as a discrete operator on the protein graph for treating diffusion, information flow and allosteric interaction problems is an exciting avenue that awaits future exploration.

In conclusion, we showed that protein graph Laplacians with harmonic residue-residue interactions lead to distributions that appear to be universal across all proteins in the non-homologous data set of 2000 crystal structures. No adjustable parameters are used in determining the shapes of the three distributions of (i) the full spectrum of density of states, (ii) the B-factors, and (iii) the entropy decrease per residue upon folding. Therefore, the outcome of this work is expected to serve as a general template for more sophisticated treatments of proteins.